# A multicaloric material as a link between electrocaloric and magnetocaloric refrigeration

Hana Ursic[1]*, Vid Bobnar[1], Barbara Malic[1], Cene Filipic[1], Marko Vrabelj[1], Silvo Drnovsek[1], Younghun Jo[2], Magdalena Wencka[3], Zdravko Kutnjak[1]

[1]Jožef Stefan Institute, Jamova cesta 39, 1000 Ljubljana, Slovenia

[2]Spin Engineering Physics Team, Korea Basic Science Institute, Daejeon 34133, Korea

[3]Institute of Molecular Physics, Polish Academy of Sciences, ul. Smoluchowskiego 17, 61-179 Poznań, Poland

*Contact author: hana.ursic@ijs.si (Hana Uršič)

Jozef Stefan Institute, Jamova Cesta 39

Ljubljana, Slovenia, 1000

http://www.nature.com/srep/journal-policies/editorial-policies/#confidentiality

**The existence and feasibility of the multicaloric, polycrystalline material $0.8Pb(Fe_{1/2}Nb_{1/2})O_3$-$0.2Pb(Mg_{1/2}W_{1/2})O_3$, exhibiting magnetocaloric and electrocaloric properties, are demonstrated. Both the electrocaloric and magnetocaloric effects are observed over a broad temperature range below room temperature. The maximum magnetocaloric temperature change of ~0.26 K is obtained with a magnetic-field amplitude of 70 kOe at a temperature of 5 K, while the maximum electrocaloric temperature change of ~0.25 K is obtained with an electric-field amplitude of 60 kV/cm at a temperature of 180 K. The material allows a multicaloric cooling mode or a separate caloric-modes operation depending on the origin of the external field and the temperature at which the field is applied.**

**Introduction**

The search for caloric materials to be applied in solid-state refrigeration has recently become one of the most active fields in condensed-matter science.[1-6] The caloric effect is related to a change of the material's entropy under the sudden application of an external field: magnetic, electric, or mechanical.[1, 2, 7, 8] Depending on the origin of the entropy change, the caloric effects can be classified as magnetocaloric (MC), electrocaloric (EC) and mechanocaloric (mC), the last of which includes the elastocaloric and barrocaloric effects.[1] For all three individual effects prototype cooling devices have already been proposed.[7, 9-11]

Lately, materials exhibiting multicaloric properties have become the "holy grail" of developments in new, solid-state cooling technologies. Until recently, the coexistence of the MC and EC effects had only been proposed theoretically.[12-16] According to a recent review of caloric materials[1] this coexistence has not been experimentally confirmed, as it is difficult to find a multiferroic material that exhibits both ferromagnetism and ferroelectricity. More recently, a study appeared reporting the multicaloric $Y_2CoMnO_6$[17]; however, as the authors explained, this material is an improper ferroelectric and for such materials the conventional methods for an EC determination are not suitable. Furthermore, in a recent publication[18] it was again stated that the multicaloric effect in a single-phase material is still awaiting an experimental confirmation. Until now, the effect has only been examined in ferromagnetic and ferroelectric composite materials.[18] In our work we show experimentally that the single-phase relaxor $0.8Pb(Fe_{1/2}Nb_{1/2})O_3$-$0.2Pb(Mg_{1/2}W_{1/2})O_3$ (PFN-PMW) exhibits both magneto- and electrocaloric effects, making it a multicaloric material.

Relaxor ferroelectrics are structurally disordered polar materials, which are characterized by both site and charge disorders and the presence of random fields. They represent a different low-temperature state of polar dielectrics, which can be regarded as an intermediate state between dipolar glasses and normal ferroelectrics.[19] PFN-PMW is a perovskite solid solution between the multiferroic PFN (ferroelectric at room temperature, becoming antiferromagnetic at low temperatures) and antiferroelectric PMW, which is diamagnetic. It has been shown that PFN-PMW exhibits a typical relaxor behavior not only in electrical, but also in magnetic properties: (i) a broad frequency dispersion in both, the electrical and magnetic susceptibilities, and (ii) a glass-like slowing down of the electric and magnetic dynamics, both following the Vogel-Fulcher behavior.[20] This means that in zero electric/magnetic fields no long-range ferroic state is established down to the lowest temperatures, and that the system is characterized by the presence of nanosized electrical and magnetic clusters of variable sizes.[21] We show in this investigation that these coexisting spin and dipolar subsystems can, with the application of external conjugate fields, lead to a multicaloric response. PFN-PMW is thus presented as a system in which both electrocaloric and magnetocaloric effects coexist or can be separately switched on or off using an external electric or magnetic field.

**Results**

The magnetization vs. magnetic field (*M-H*) hysteresis loop of the PFN-PMW material at 5 K is shown in **Figure 1a**. Note that the density of the sintered pellets is 8.15 g/cm$^3$. When increasing the temperature (inset, **Figure 1a**), the hysteresis loops become suppressed. The temperature dependence of the magnetization is shown in **Figure 1a**, where a temperature increase results in a decrease of the magnetization. The highest measured magnetization of 2.3 emu/g is achieved at 5 K and 70 kOe. In **Figure 1b** the magnetocaloric temperature changes $ΔT_{MC}$ vs. $T$ and $H$ are shown. The MC temperature change $ΔT_{MC}$ is 0.26 K at 5 K and 70 kOe, and it gradually decreases with an increasing temperature, but it can still be detected up to 300 K. Furthermore, $ΔT_{MC}$ increases with an increasing magnetic field $H$; the largest increase in $ΔT_{MC}$ is observed at the lowest measured temperature of 5 K (inset in **Figure 1b**).

To validate the magnetic measurements on the PFN-PMW material using two independent methods, in addition to the Quantum Design Physical Property Measurement System (PPMS) measurements (**Figure 1a**), the $M$ vs. $T$ was also measured using a Superconducting Quantum Interference Device (SQUID) (**Figure 1c**, inset). The latter measurements were performed at lower magnetic fields, i.e., from 0.5 to 10 kOe; however, a similar trend was observed in the measurements made using both methods (see the comparison in **Figure 1c**).

The results of the direct and indirect EC measurements of the PFN-PMW are shown in **Figure 2**. As expected, the $ΔT_{EC}$ increases with increasing temperature as it approaches the phase transformation at ~270 K. For example, the $ΔT_{EC}$ is 0.15 K at 180 K and 40 kV/cm, and it increases to 0.21 K when the temperature increases by 20 K. The $ΔT_{EC}$ also increases with an increasing electric field. The highest $ΔT_{EC}$ of 0.245 K was measured at 60 kV/cm and 180 K. During the direct EC measurements no significant Joule heating was detected up to 220 K, although when increasing the temperature, some Joule heating (≥ 0.06 K) was observed.

Hence, we show experimentally that in the PFN-PMW material the electrocaloric and magnetocaloric effects coexist and can be separately switched on or off using an external electric or magnetic field. A schematic diagram of such electrocaloric, magnetocaloric and multicaloric cooling cycles of the studied material is shown in **Figure 3**. In spite of the fact that the $\Delta T_{MC}$ and $\Delta T_{EC}$ are relatively small (i.e., maximum measured values of ~0.26 K and ~0.25 K, respectively), the feasibility of such a single-phase multicaloric MC and EC material is proven, which should promote the further development of multicaloric materials with larger cooling responses.

**Discussion**

A missing link between magnetocaloric and electrocaloric cooling is proposed and experimentally demonstrated with the polycrystalline PFN-PMW multicaloric material. We show that such a multicaloric material exists and, furthermore, that the magnetocaloric and electrocaloric modes can be applied in two different temperature regions, extending the operating temperature range of the caloric material from 5 K up to 220 K. Since in this temperature range both caloric effects coexist, the application of both stimuli can enhance the total caloric effect. Such a combined caloric effect in multiferroic materials can lead to hybrid cooling systems of a new generation that are capable of working across a broad temperature range.

**Methods**

For the synthesis of the PFN-PMW powder, PbO (99.9 %, Sigma-Aldrich, 211907), $Fe_2O_3$ (99.9 %, Alfa, 014680-Ventron), $Nb_2O_5$ (99.9 %, Sigma-Aldrich, 208515), $WO_3$ (99.8%, Alfa, 82120-Ventron), and MgO (98 %, Sigma-Aldrich, 24338) were used. The homogenized, stoichiometric mixture (200 g) was mechanochemically activated in a high-energy planetary mill (Retsch, Model PM 400) for 40 h at 300 rpm using a tungsten carbide milling vial and balls. The synthesized powder was milled in an attrition mill with yttria-stabilized zirconia (YSZ) balls in isopropanol for 4 h at 800 rpm. The powder was then uniaxially pressed into disks and further consolidated by isostatic pressing at 300 MPa. The powder compacts were sintered in double alumina crucibles in the presence of a packing powder with the same chemical composition to avoid possible PbO losses. The compacts were sintered at 1123 K for 2 h in an oxygen atmosphere at heating and cooling rates of 2 K/min. The density was determined using a Micromeritics – AccuPyc II 1340 gas pycnometer. The X-ray diffraction pattern and the microstructure of the ceramics are summarized in Supplementary Information, **Figures S1, S2**.

The *M-H* curves were detected at temperatures between 5 K and 300 K with a 16-T PPMS using the AC Measurement System (ACMS) option. Furthermore, the magnetization vs. temperature at different magnetic fields (from 0.5 kOe to 10 kOe) was measured using a SQUID from 20 K to 270 K. The sample's weight was 30 mg.

For the electrical measurements, the faces of the disks with a diameter of 6 mm and a thickness of 70 µm were coated with Cr/Au by RF-magnetron sputtering (5 Pascal). For the direct EC measurements, a modified high-resolution calorimeter was used, since it allowed a precise temperature stabilization of the bath (± 0.1 mK). The temperature was measured with a small-bead thermistor. The direct EC measurements were supported by indirect methods in which the magnetocaloric and electrocaloric temperature changes were calculated using equations (1) and (2), given in Supporting Information. The details of the method can be found in [22, 23]. The methodology as well as the measurements of the complex dielectric constant $\varepsilon^*(\nu,T)$, the polarization-electric (*P-E*) field response and the heat capacity *Cp* versus temperature are summarized in Supplementary Information (**Figures S3-S5**).

**Acknowledgements**

The authors would like to thank B. Kmet, J. Cilenšek and the Slovenian Research Agency for financial support (programs P2-0105, P1-0125; projects L2-6768, J2-7526). V.B. is grateful for the hospitality shown to him during his stay at the Korea Basic Science Institute in the frame of the European program Marie Curie Actions, FP7-PEOPLE-2011-IRSES, NanoMag, 295190, (2012). MW and ZK acknowledge to Polish Academy of Sciences and Slovenian Academy of Sciences and Arts for financial contribution (joint research project 2015-2017 "Multicaloric relaxor materials for new cooling technologies"). Authors thank Z. Jaglicic for his help in performing SQUID measurements.

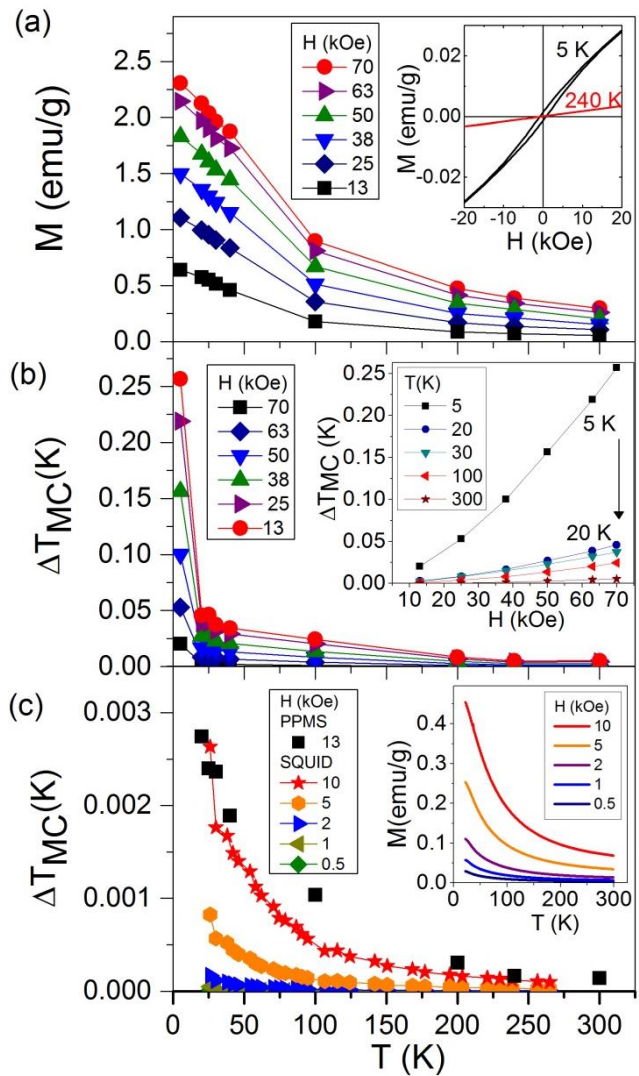

**Figure 1:** (a) *M* vs. *T* at *H* from 13 to 70 kOe measured using a Quantum Design PPMS. Lines are a guide for the eye. Inset: *M-H* hysteresis loops measured at 5 and 240 K. (b) *ΔT$_{MC}$* vs. *T* and in inset *ΔT$_{MC}$* vs. *H* calculated from the measurements given in (a). The black arrow indicates the decrease in temperature. (c) *ΔT$_{MC}$* vs. *T* calculated from *M-T* measurements measured using a SQUID and shown in the inset. For comparison the *ΔT$_{MC}$* vs. *T* measured using a Quantum Design PPMS at 13 kOe is also given (black squares).

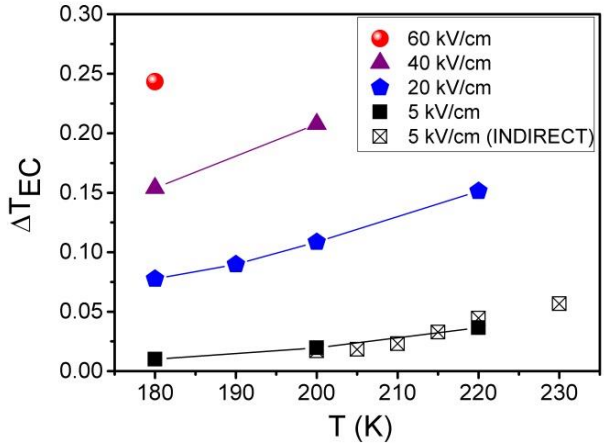

**Figure 2:** *ΔT$_{EC}$* vs. *T* at electric field amplitudes from 5 to 60 kV/cm. The solid and crossed squares represent the direct and indirect EC measurements, respectively. Lines are a guide for the eye.

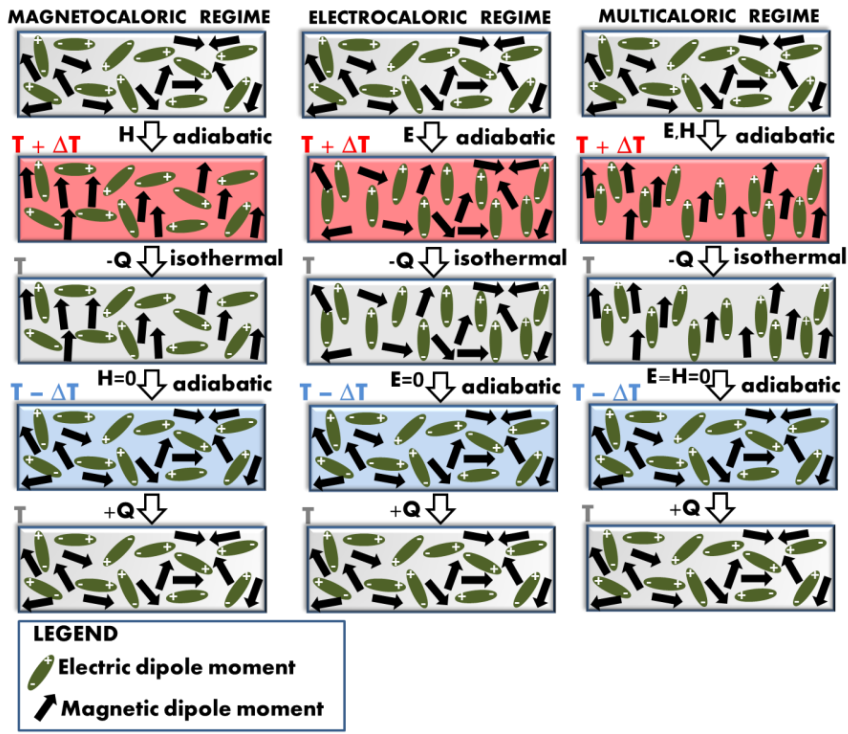

**Figure 3:** A schematic diagram of the magnetocaloric (left), electrocaloric (middle) and multicaloric (right) cooling cycles.